\documentclass[a4paper,11pt]{article}   
\usepackage{authblk} 
\usepackage[top=2cm,bottom=2cm,left=2cm,right=2cm]{geometry}

\usepackage{listings}
\usepackage{color}

\usepackage{xspace}
\usepackage{tikz}
\usepackage{caption}
\usepackage{subcaption}
\usetikzlibrary{mindmap}

\newcommand{\progname}{\textsf}
\newcommand{\compname}{\textsf} 

\newcommand{\lrs}{\progname{lrs}\xspace}

\newcommand{\mplrs}{\progname{mplrs}\xspace}
\newcommand{\mts}{\progname{mts}\xspace}

\newcommand{\btree}{\progname{btree}\xspace}
\newcommand{\mtree}{\progname{mtree}\xspace}

\newcommand{\btop}{\progname{btopsorts}\xspace}
\newcommand{\mtop}{\progname{mtopsorts}\xspace}
\newcommand{\vr}{\progname{VR}\xspace}
\newcommand{\genle}{\progname{Genle}\xspace}
\newcommand{\gray}{\progname{grayspan}\xspace}
\newcommand{\graysp}{\progname{grayspspan}\xspace}

\newcommand{\maic}{\compname{mai32}\xspace}

\newcommand{\mainew}{\compname{mai32abcd}\xspace}
\newcommand{\mainewer}{\compname{mai32ef}\xspace}

\newcommand{\gnuplot}{\progname{gnuplot}\xspace}

\newcommand{\Adj}{\textrm{Adj}}
\newcommand{\mydepth}{\ensuremath{\mathit{depth}}\xspace}
\newcommand{\mymaxdepth}{\ensuremath{\mathit{max\_depth}}\xspace}
\newcommand{\startvertex}{\ensuremath{\mathit{start\_vertex}}\xspace}
\newcommand{\maxnodes}{\ensuremath{\mathit{max\_nodes}}\xspace}
\newcommand{\myfalse}{\ensuremath{\mathbf{false}}\xspace}
\newcommand{\mytrue}{\ensuremath{\mathbf{true}}\xspace}
\newcommand{\mycount}{\ensuremath{\mathit{count}}\xspace}
\newcommand{\putoutput}{\ensuremath{\textrm{put\_output}}\xspace}
\newcommand{\unexplored}{\ensuremath{\mathit{unexplored}}\xspace}
\newcommand{\lmin}{\ensuremath{\mathit{lmin}}\xspace}
\newcommand{\lmax}{\ensuremath{\mathit{lmax}}\xspace}
\newcommand{\myscale}{\ensuremath{\mathit{scale}}\xspace}
\newcommand{\maxbuf}{\ensuremath{\mathit{maxbuf}}\xspace}

\usepackage{todonotes}
\usepackage{amsmath}
\usepackage{amssymb}
\usepackage{latexsym}
\usepackage{graphicx}
\usepackage{color}
\usepackage{hyperref}
\usepackage{slashbox}
\usepackage{algorithm}
\usepackage{algpseudocode}

\definecolor{darkblue}{rgb}{0,0,0.6}

\include{./mathlib}
\include{./macros}

\begin{document}

\lstset{language=C, xleftmargin=\parindent, showspaces=false, showstringspaces=false} 

\title{A parallel framework for reverse search using \mts\thanks{This work was partially supported by JSPS 
Kakenhi Grants 16H02785, 23700019 and 15H00847, Grant-in-Aid for Scientific Research 
on Innovative Areas, `Exploring the Limits of Computation (ELC)'.}}

\author[1]{David Avis}
\author[2]{Charles Jordan}
\affil[1]{School of Informatics, Kyoto University, Kyoto, Japan and 
          School of Computer Science,
          McGill University, Montr{\'e}al, Qu{\'e}bec, Canada\\
          \texttt{avis@cs.mcgill.ca}}
\affil[2]{Graduate School of Information Science and Technology, 
          Hokkaido University, Japan\\
          \texttt{skip@ist.hokudai.ac.jp}}

\maketitle

\begin{abstract}
We describe \mts, which is a generic framework for 
parallelizing certain types of tree search programs, that
(a) provides a single common wrapper containing all
of the parallelization, and (b) minimizes the changes needed to the
existing single processor legacy code.
The \mts code was derived
from ideas used to develop \mplrs, a parallelization of the reverse search vertex
enumeration code \lrs. The tree search properties required for the use of \mts
are satisfied by any reverse search algorithm as well as other tree search methods
such as backtracking and branch and bound. 
\mts is programmed in C, uses the MPI parallel environment, and can be run on a network of computers. 

As examples we parallelize two simple existing reverse search codes: generating topological
orderings and generating spanning trees of a graph. We give computational results comparing
the parallel codes with state of the art sequential codes for the same problems.

\noindent{}Keywords: reverse search, parallel processing, topological orderings, spanning trees\\
Mathematics Subject Classification (2000) 90C05
\end{abstract}

\section{Introduction}
\label{sec:intro}
Although multicore hardware is now pervasive there are relatively few legacy codes
that can profit from it. In some cases the underlying algorithm seems inherently sequential,
rendering parallelization difficult if not impossible. In other cases the underlying algorithm
may in principle be easy to parallelize. Nevertheless, existing 
widely-used and debugged sequential codes
may be extremely complex and adding parallelization directly into such codes can
be difficult and error prone. 
A simpler approach would be to design a wrapper that provides the parallelization
and hopefully requires very little modification of the existing code.
This approach was successfully applied by the authors to 
the vertex enumeration code \lrs, resulting in the parallel \mplrs code~\cite{AJ15b}.
The ideas used in \mplrs can be used for a wide variety of tree search algorithms
including reverse search, backtracking, branch and bound, etc.
This motivated us to develop a generic wrapper, \mts, for parallelizing them.
In this note we explain how to modify a reverse search code to enable parallelization
by \mts.
We then describe in detail how to parallelize two simple existing reverse search codes.
Finally, we give computational results showing nearly linear speedup on a cluster with 
192 cores.
\section{Background}
\label{sec:back}

Reverse search is a technique for generating large, relatively unstructured, sets of discrete
objects~\cite{AF93}. 
Some simple C implementations were given in
the tutorial~\cite{tutorial} which we will extend in this note to
allow parallel processing via the \mts package.
We first give an outline of reverse search and the necessary
modifications required for parallelization.
This description is essentially the one given in~\cite{AJ15b}.

In its most basic form, reverse search can be viewed as the traversal of a spanning tree, called the reverse
search tree $T$, of a graph $G=(V,E)$ whose nodes are the objects to be generated. Edges in the graph are
specified by an adjacency oracle, and the subset of edges of the reverse search tree are
determined by an auxiliary function, which can be thought of as a local search function $f$ for an
optimization problem defined on the set of objects to be generated. One vertex, $v^*$, is designated
as the {\em target} vertex. For every other vertex $v \in V$ 
repeated application of $f$ must generate a
path in $G$ from $v$ to $v^*$. The set of these paths defines the reverse search tree $T$, which has root $v^*$.

A reverse search is initiated at $v^*$, and only edges of the reverse search tree are traversed.
When a node is visited, the corresponding object is output.  Since there is no possibility of
visiting a node by different paths, the visited nodes do not
need to be stored.  Backtracking can be performed in the
standard way using a stack, but this is not required as the local search function can be used for
this purpose. This implies two critical features that are useful for effective parallelization.
Firstly, it is not necessary to store more than one node of the tree at any
given time and no database is required for visited nodes. 
Secondly, it is possible to {\em restart} the enumeration process from
any given node in the tree using only a description of this one node.

In the basic setting described here a few properties are required. Firstly, the
underlying graph $G$ must be connected and an upper bound on the maximum vertex degree, $\Delta$, must
be known.  The performance of the method depends on $G$ having $\Delta$ as low as
possible.  The adjacency oracle must be capable of generating the adjacent vertices of some given
vertex $v$ sequentially and without repetition.  This  is  done  by  specifying  a  function  
$\Adj(v,j)$, where $v$ is  a  vertex  of $G$ and $j = 1,2,\ldots,\Delta$.  Each value of $\Adj(v, j)$ is
either a vertex adjacent to $v$ or null. Each vertex adjacent to $v$ appears precisely once as $j$ ranges
over its possible values.  For each vertex $v \neq v^*$
the local search function $f(v)$ returns the tuple $(u,j)$ where $v = \Adj(u,j)$ such that $u$
is $v$'s parent in $T$.
Pseudocode is given in Algorithm~\ref{rsalg1} and C implementations for
several simple enumeration problems are given at~\cite{tutorial}.
For convenience later, we do not output the root vertex $v^*$ in the pseudocode shown.
Note that the vertices are output as a continuous stream.

\begin{algorithm}
\begin{algorithmic}[1]
\Procedure{rs}{$v^*$, $\Delta$, $\Adj$, $f$}
        \State $v \gets v^*$~~~$j \gets 0$~~~$\mydepth \gets 0 $
        \Repeat
       	\While {$j < \Delta$}
                \State $j \gets j+1$
		\If {$f(\Adj(v,j)) = v$}  \Comment{forward step}
			\State $v \gets \Adj(v,j)~~~~~$  
			\State $j \gets 0$ 
                        \State $\mydepth \gets \mydepth+1$         
			\State {\bf output $v$}         
                \EndIf
        \EndWhile
        \If {$\mydepth > 0$}   \Comment{backtrack step}                 
		\State $(v,j) \gets f(v)$
                \State $\mydepth \gets \mydepth-1  $       
        \EndIf
        \Until {$\mydepth=0$ {\bf and} $j=\Delta$}
\EndProcedure
\end{algorithmic}
\caption{Generic Reverse Search}
\label{rsalg1}
\end{algorithm}
Observe that Algorithm~\ref{rsalg1} does not 
require the parameter  $v^*$ to be the root of the search tree. If
an arbitrary node in the tree is given, the algorithm reports the subtree
rooted at this node and terminates. This is the key property that
allows parallelization, as we discuss next. 

\section{Maximum depth and budgeting}
\label{sec:budg}

In order to parallelize Algorithm~\ref{rsalg1} we need to break up the search
of the entire tree $T$
into searching a set of subtrees.
We do this in two ways: by limiting the search depth and by limiting the number
of nodes visited. In both cases, unexplored subtrees are generated and these
must be flagged for later processing.
To implement this we will supply three additional parameters: 
\begin{itemize}
\item
\startvertex is the vertex from which the reverse search should be initiated and replaces $v^*$.
\item
\mymaxdepth is the depth at which forward steps are terminated.
\item
\maxnodes is the number of nodes to generate before terminating and reporting unexplored subtrees.
\end{itemize}
Both \mymaxdepth and \maxnodes are assumed to be positive, for otherwise there is no work to do.
The modified algorithm is shown in Algorithm~\ref{bts}. 

\begin{algorithm}[htb]
\begin{algorithmic}[1]
\Procedure{bts}{$\startvertex$, $\Delta$, $\Adj$, $f$, $\mymaxdepth$, $\maxnodes$}
        \State $j \gets 0~~~v \gets \startvertex~~~\mycount \gets 0 ~~~\mydepth \gets 0$
        \Repeat
        \State $\unexplored \gets \myfalse$
        \While {$j < \Delta$ {\bf and} $\unexplored = \myfalse$ }
                \State $j \gets j+1$
                \If {$f(\Adj(v,j)) = v$}  \Comment{forward step}
                        \State $v \gets \Adj(v,j)~~~~~$
                        \State $j \gets 0$ 
                        \State $\mycount \gets \mycount+1$
                        \State $\mydepth \gets \mydepth + 1$
                        \If {$\mycount \ge \maxnodes$ {\bf or} $\mydepth = \mymaxdepth$} \Comment{budget is exhausted}
                            \State $\unexplored \gets \mytrue$
                        \EndIf
                        \State \putoutput $(v,\unexplored)$
                \EndIf
        \EndWhile
        \If {$\mydepth > 0$}   \Comment{backtrack step}
                \State $(v,j) \gets f(v)$
                \State $\mydepth \gets \mydepth - 1$
        \EndIf
        \Until {$\mydepth = 0$ {\bf and} $j=\Delta$}
\EndProcedure
\end{algorithmic}
\caption{Budgeted Reverse Search}
\label{bts}
\end{algorithm}
Comparing Algorithm~\ref{rsalg1} and Algorithm~\ref{bts}, we note several changes. 
Firstly an integer variable \mycount is introduced to keep track
of how many tree nodes have been visited, a process we call {\em budgeting}. 
Secondly, a flag \unexplored is introduced to distinguish the tree nodes whose subtrees
have not been explored. It is initialized as {\em false} on line 4.
The flag is set to \mytrue in line 13 if either 
the budget of \maxnodes has been exhausted
or a depth of \mymaxdepth has been reached. 
Each node encountered on a forward step is output via the routine \putoutput
on line 15.
In single processor mode the output is simply sent to the output file with a flag added
to unexplored nodes. In multi-processor mode, the output is synchronized and
unexplored nodes are returned to the controlling master process.

Backtracking is as in Algorithm~\ref{rsalg1}.
After each backtrack step the \unexplored flag is set to \myfalse in line 4.
If the budget constraint has been exhausted then \unexplored will again be set
to \mytrue in line 13 after the first forward step.
In this way all unexplored siblings of nodes on the backtrack path to the root are flagged
in \putoutput.
If the budget is not exhausted then forward steps continue until either it is, \mymaxdepth is reached, or we reach a leaf.

To output all nodes in the subtree of $T$ rooted at $v$ we set  
$\startvertex=v$, $\maxnodes=+\infty$ and $\mymaxdepth=+\infty$.
This reduces to Algorithm~\ref{rsalg1} if $v=v^*$.
To break up $T$ into subtrees we have two options that can be combined.
Firstly we can set the \mymaxdepth parameter resulting in all nodes at
that depth to be flagged as unexplored.
Secondly we can set the budget parameter \maxnodes.
In this case, once this many nodes have been explored the current node
and all unexplored siblings on the backtrack path to the root are output
and flagged as unexplored.

\begin{figure}[htbp]
\centering
\begin{tikzpicture}[grow cyclic, align=flush center,
    level 1/.style={level distance=3cm,sibling angle=90},
    level 2/.style={level distance=1.6cm,sibling angle=45}]
\node{$0$}
 child { node {$1$}
         child { node {$2$} }
         child { node {$3$} }
         child { node {$4$} }
         child { node {$5$} }
         child { node {$6$} }
       }
 child { node {$7$}
         child { node {$8$} }
         child { node {$9$} }
         child { node {$10$} }
         child { node {$11$}
             child { node {$12$} }
             child { node {$13$}
                   child { node {$14$} }
                   }
              }
         child { node {$15$} }
         child { node {$16$}
              child { node {$17$} }
              }
       }
 child { node {$18$}
         child { node {$19$} }
         child { node {$20$ } }
         child { node {$21$} }
       }
 child { node {$22$}
         child { node {$23$} }
         child { node {$24$} }
       };
\end{tikzpicture}
\caption{Tree with 25 nodes and $\Delta=6$}
\label{fig:example}
\end{figure}
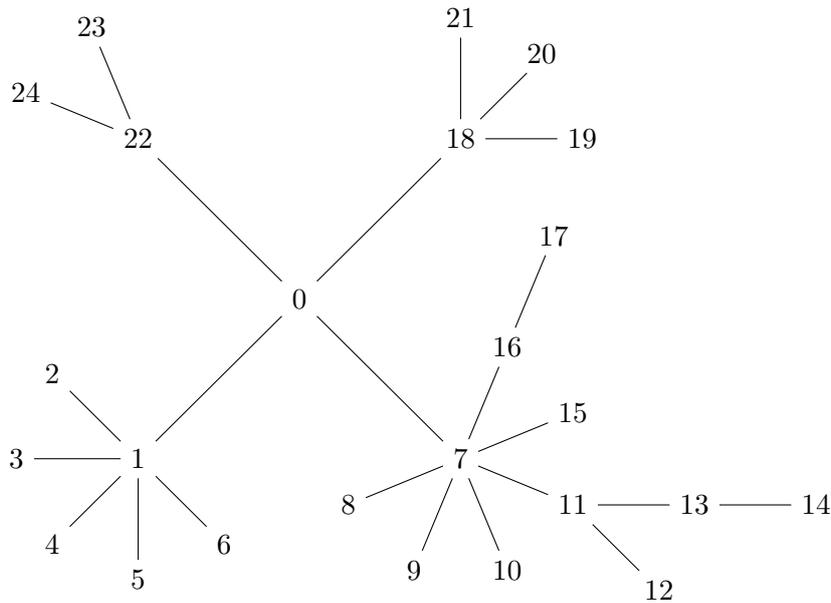

Consider the tree in Figure~\ref{fig:example} which has 25 nodes, $\Delta=6$ and is
rooted at vertex 0.
For convenience the nodes are numbered 0,1,\ldots,24 in reverse search order
but this is in no way essential.
If we set  $\mymaxdepth=1$ and $\maxnodes=+\infty$ then only nodes 1,7,18,22 are visited 
and output
with $\unexplored=\mytrue$.
Now suppose we set the parameter $\maxnodes=13$ and $\mymaxdepth=+\infty$.
Firstly nodes 1,\ldots,12 are output
with $\unexplored=\myfalse$.
Then, nodes 13,15,16,18,22 are output with $\unexplored=\mytrue$.
Alternatively, if we set $\maxnodes=8$ then nodes 1,2,\ldots,7 are output with $\unexplored=\myfalse$
and nodes 8,9,10,11,15,16,18,22 are output with $\unexplored=\mytrue$.

\section{An example \mts interface: topsorts}
\label{sec:mts}
In the tutorial~\cite{tutorial} a C implementation ({\em per.c}) is given
for the reverse search algorithm for generating permutations.
A small modification of this code generates all
linear extensions of a partially ordered set that is given by
a directed acyclic graph (DAG). Such linear extensions are also called
topological sorts or topological orderings.
The code modification is given as Exercise 5.1 and a solution to
the exercise 
({\em topsorts.c}) appears at the URL~\cite{tutorial}.
In this section we describe in detail how to modify this code to allow
parallelization via the \mts interface.

It is convenient to describe the procedure as two phases. Phase
1 implements max-depth and budgeting and organizes the internal data in a suitable
way. This involves modifying an implementation
of Algorithm~\ref{rsalg1} to an implementation of Algorithm~\ref{bts}
that can be independently tested. We also prepare a global data structure  bts\_data
which contains problem data obtained from the input.
In Phase 2 we build a node structure for use by the \mts wrapper and add necessary
routines to allow initialization and I/O in a parallel setting. In practice
this involves sharing a common header file with \mts. The resulting program
can be compiled as a stand-alone code or as a parallel code with no change in the 
source files.
The code shown in the two subsections below is essentially that as given at
\cite{tutorial} with a few nonessential parts deleted.

\subsection{Phase 1 : atopsorts.c}
\begin{itemize}
\item
    Internal data grouped in bts\_data structure
{\footnotesize
\begin{lstlisting}
struct bts_data {
        int n;                 /* number of nodes and edges in graph */
        int m;
        int A[100][100];       /* the graph */
        int countonly;         /* TRUE means nodes are counted but not output */
};

\end{lstlisting}
}

\item
    Initialization of bts\_data from the input file via {\em bts\_init}:
{\footnotesize
\begin{lstlisting}
bts_data *bts_init(int argc, char **argv)
{
  int i,j,k,m,n;
  bts_data *b = malloc(sizeof(bts_data));
  b->countonly=FALSE;

/* process command line arguments         */
        for (i=1; i<argc; i++)
          if (strcmp(argv[i],"-countonly") == 0)
             b->countonly=TRUE;

/* read input file and build bts_data     */
  scanf("%d %d", &b->n, &b->m);
  n=b->n;
  m=b->m;
  for (i=1; i<=n; i++)
     for (j=1; j<=n; j++)
       b->A[i][j]=0;

  for (k=1; k<=m; k++)
    {
     scanf("%d %d",&i,&j);
     b->A[i][j]=1;
    }
        return b;
}
\end{lstlisting}
}

\item
    \maxnodes and \mymaxdepth parameters are introduced to the reverse
search code as described in Algorithm~\ref{bts}:
{\footnotesize
\begin{lstlisting}
long bts(bts_data *b, perm v, long max_depth, long max_nodes)
{
  int j=0, depth=0, unexplored;
  long count=0;
  int n = b->n;
  int maxdeg=n-1;
  do
  {
      unexplored = FALSE;
      while (j < maxdeg && !unexplored)
      {
         j++;
         if (reverse(b, v, j))
          {                        /* forward step */
             Adj(v, j);
             depth++;
             count++;
             if (count >= max_nodes || depth == max_depth)
                  unexplored=TRUE;
             put_output(b,v,depth,unexplored);
             j = 0;
          }
      }   /* end while */

     if(depth > 0)
          {                        /* backtrack step */
             j = backtrack(b, v);
             depth--;
          }
 } while(depth > 0 || j < maxdeg);
  return count;
}


\end{lstlisting}
}

\item
    {\em main()} is modified to call {\em bts\_init}, process command line arguments to test
budgeting and call {\em bts()} to perform the tree search:

{\footnotesize
\begin{lstlisting}
int main(int argc, char **argv)
/* generate topsorts with budgeting and bts_data structure */
{
  bts_data *b;
  perm v;
  long count;
  int i,n;
  long max_nodes=LONG_MAX, max_depth=LONG_MAX;

  b=bts_init(argc,argv);
  n=b->n;

  for (i=1; i<=n; i++)
    v[i] = i;

  for (i=1; i<argc; i++)
    {
          if (strcmp(argv[i],"-maxd") == 0)
                max_depth = atoi(argv[i+1]);
          if (strcmp(argv[i],"-maxnodes") == 0)
                 max_nodes = atoi(argv[i+1]);
    }

/* the tree search is now initiated from root */
  put_output(b,v,0,0);     /* first output */
  count = bts(b,v,max_depth,max_nodes);

/* finish up */
  printf("\nnumber of permutations=%ld\n", count+1); /* count the root */

  free(b);
  return 0;
}
\end{lstlisting}
}
\end{itemize}

\subsection{Phase 2 : btopsorts.c}
\label{subsec:btopphase2}
In the second phase we add the `hooks' that allow communication with \mts.
This involves defining a Node structure which holds all necessary information
about a node in the search tree. The roots of unexplored subtrees are maintained
by \mts for parallel processing. Therefore whenever a search terminates due
to the \maxnodes or \mymaxdepth restrictions, the Node structure of each unexplored
tree node is returned to \mts. As we do not wish to customize \mts for
each application, we use a very generic node structure. The user should pack
and unpack the necessary data into this structure as required. The Node
structure is defined in the {\em mts.h} header.
We now give some more specific details.
Most of the code is common to single and multithread mode. The few
exceptional cases are handled by
the compile switch {\em \#ifdef MTS}.
\begin{itemize}
\item
    The generic Node data structure in {\em mts.h} is:
{\footnotesize
\begin{lstlisting}
typedef struct  node_v {
        long *vlong;            /* tree node longs (for user) */
        unsigned int size_vlong;/* number of longs in vlong */
        int *vint;              /* tree node ints (for user)  */
        unsigned int size_vint; /* number of ints in vint */
        char *vchar;            /* tree node chars (for user) */
        unsigned int size_vchar;/* number of chars in vchar */
        float *vfloat;          /* tree node floats (for user) */
        unsigned int size_vfloat;/*number of floats in vfloat */

        int unexplored; /* this subtree is unexplored */
        long depth;     /* depth of v in tree */
        struct node_v *next;    /* for list of unexplored nodes */
} Node;
\end{lstlisting}
}
This is not user modifiable without changing \mts.c itself. 
\item
    The tree search begins from a single root node, which is allocated and
initialized in:

{\footnotesize
\begin{lstlisting}
Node *get_root(bts_data *b)
/* return a node which is the root of the search tree  */
{
   int i,n;
   Node *root = malloc(sizeof(Node));

   n=b->n;
   root->vlong = malloc(sizeof(long)*(n+1));
   root->size_vlong = n+1;
   root->vint = NULL; root->size_vint =0;
   root->vchar = NULL; root->size_vchar = 0;
   root->vfloat = NULL; root->size_vfloat = 0;
   root->depth = 0;
   root->unexplored = FALSE;

   for (i=1; i<=n; i++)  /* root permutation for topsort is 1,2,...,n */
            root->vlong[i] = i;

   if (b->countonly == FALSE)
        put_output(b,root);  /* first output ! */
   return root;
}
\end{lstlisting}
}
In {\em btopsorts.c} a call to {\em get\_root()} is made immediately after the call to {\em bts\_init()}.
When using \mts, a call is made to {\em get\_root()} as part of the 
initialization process.  This is called only once, \mts deals with transferring
nodes between processors.
\item
Options for {\em btopsorts.c} are collected in the structure: 
{\footnotesize
\begin{lstlisting}
const mtsoption bts_options[] = {
 {"-countonly", 0},     /* -countonly option has no parameters */
 {"-prune", 1},         /* -prune option has one parameter */
};
const int bts_options_size = 2;
\end{lstlisting}
}
which has currently two user defined options.
The first option, {\em -countonly}, takes no parameters and the second, {\em -prune} (described in Section~\ref{extras}), 
takes one. 
Additional options can be added here as long as bts\_options\_size is updated 
and the options do not
conflict with the \mts option list (described in Section~\ref{subsec:mtsopts}).
\item
The input file is handled by \mts which converts it to a string pointed to by file pointer $f$. The
user simply needs to convert all {\em scanf(\ldots)} statements to
{\em fscanf(f, \ldots)}.
{\em mtslib.c} contains {\em get\_input()} to convert input to Data for internal use.
No user modification is required or recommended.

Output is handled by \mts in multiprocessor mode. Output to stdout and stderr
is controlled by the values set for b$\rightarrow$output and b$\rightarrow$err in {\em bts\_init()}.
This is transparent to the user, who needs
only replace all instances of {\em printf(\ldots)} by {\em tsprintf(b$\rightarrow$output,\ldots)}
and also instances of {\em fprintf(stderr,\ldots)} by {\em tsprintf(b$\rightarrow$err,\ldots)}.

These I/O handling considerations require a slight modification of
{\em bts\_data} and associated declarations:
Note in particular the extra two lines required in the definition of {\em bts\_data}
and the compile switch.

{\footnotesize
\begin{lstlisting}
#ifdef MTS
#define tsprintf stream_printf
#define tstream mts_stream
#else
#define tsprintf fprintf
#define tstream FILE
#endif

typedef long Perm[100];

/* bts_data is built from the input and can be user modified  */
/* except where mentioned                                     */

struct bts_data {
        int n;                 /* number of nodes and edges in graph */
        int m;
        int A[100][100];       /* the graph */
        int countonly;         /* TRUE means nodes are counted but not output */
        int prune;             /* mark unexplored only if nchild > prune */
        tstream *output;       /* output stream for stdout */
        tstream *err;          /* output stream for stderr */
};
\end{lstlisting}
}

\item
A few small changes to {\em put\_output()}, making use of the compile switch,
are required
to be sure the output goes in the right place:
{\footnotesize
\begin{lstlisting}
void put_output(bts_data *b, Node *treenode)
{
 int i;
 if(!b->countonly)
 {
   for (i=1; i<=b->n; i++)
      tsprintf(b->output," %ld", treenode->vlong[i]);
   tsprintf(b->output, " d=%ld",treenode->depth);
#ifndef MTS
   if (treenode->unexplored)
       tsprintf(b->output, " *unexplored");
#endif
   tsprintf(b->output, "\n");
 }
#ifdef MTS
 if (treenode->unexplored)
   return_unexp(treenode);
#endif
}
\end{lstlisting}
}
The {\em return\_unexp()} routine is supplied in {\em mts.c}.
\item
A few small changes are required to {\em bts\_init()}. Omitting
code already given above, this becomes: 
{\footnotesize
\begin{lstlisting}
bts_data *bts_init(int argc, char **argv, Data *in, int proc_no)
{
        int i, j, k, n,m;
        FILE *f;

        bts_data *b = malloc(sizeof(bts_data));
        b->countonly=FALSE;

/* process command line arguments to btopsorts - as above, omitted */

        f = open_string(in->input);
#ifdef MTS
        b->output = open_stream(MTSOUT);
        b->err = open_stream(MTSERR);
#else
        b->output = stdout;
        b->err = stderr;
#endif
        i = fscanf(f, "%d %d\n", &b->n, &b->m);
        if (i == EOF )
        {
           tsprintf(b->err, "no input found\n");
           fclose(f);
#ifdef MTS
           close_string();
#endif
           return NULL;
        }

        n=b->n;
        m=b->m;

/* read input data containing the input graph - as above, omitted */

        return b;
}                      
\end{lstlisting}
}
\item
Finally there is {\em bts\_finish()} which is called by the master program once
after all worker processors have terminated. This routine is not required for reverse search
applications and is included only for future compatibility.
It is intended to be used by applications that use shared\_data, such as 
branch and bound, to provide
a final solution to a problem.

{\footnotesize
\begin{lstlisting}
void bts_finish(bts_data *bdata, shared_data *sdata, int size, int checkpointing)
{
  return;
}
\end{lstlisting}
}

\end{itemize}
By itself {\em btopsorts.c} compiles to the standalone program \btop, but it
becomes the parallel code \mtop if compiled with the 
{\em mts.c} wrapper using the MTS flag.

\subsection{Extras}
\label{extras}
There are some additional features implemented in \mts that are not essential
but may prove useful.
They are documented in the code available at~\cite{tutorial} and include:
\begin{itemize}
\item
{\em cleanstop()} and {\em emergencystop()}.
These allow a user process to cleanly close down the entire parallel execution.
\item
Shared data. This allows user processes to pass data back to \mts
which may be useful for other processes. This is not required for reverse
search applications but will be useful for applications such as branch and bound,
satisfiability, game tree search, etc.
\item
Output blocks. Although \mts guarantees that output from each call to
{\em stream\_printf} is synchronized with other processes, it may be
desirable to synchronize a block of {\em stream\_printf} output.  If
the application opens an \emph{output block}, all output produced until
the application closes this \emph{output block} (or \emph{bts} returns) will
be printed as a block without interruption from other workers.
\item
Pruning. The efficiency of \mts depends on keeping the job list non-empty
until the end of the computation, without letting it get too large. Depending on
the application, there may be a substantial restart cost for each unexplored
subtree. Surely there is no need to return a leaf as an unexplored node, and the
{\em prune=0} option checks for this. Further, if an unexplored node has only
one child it may be advantageous to explore further, terminating either at
a leaf or at a node with two or more children, which is returned
as {\em unexplored}. The {\em prune=1}
option handles this condition, meaning that no isolated nodes or paths are
returned as unexplored.
Note that pruning is not a built-in \mts option; it is an example option
that applications may wish to include.  An example of pruning is implemented
in \mtop, as mentioned in Section~\ref{subsec:btopphase2}.

\end{itemize} 

\subsection{A second example : spanning trees}
\label{sec:tree}
In the tutorial~\cite{tutorial} a C implementation ({\em tree.c}) is given       
for the reverse search algorithm for all spanning trees of the complete graph.
An extension of this to generate all spanning trees of 
a given graph is stated as Exercise 6.3.
Applying Phase 1 and 2 as described above results in the codes {\em atree.c}
and {\em btree.c}. The \mts wrapper may be directly compiled with {\em btree.c}
to provide the parallel implementation \mtree.
All of these codes are given at the URL~\cite{tutorial}.

\section{Using the \mts interface}
\label{subsec:mtsopts}

\subsection{Building \mts}
Applications may have additional requirements, but the
requirements for \mts are fairly simple.  The most important
is an MPI implementation.  We develop and test on Open MPI and Intel MPI,
but any reasonably modern implementation should work.  We use the common
\progname{mpicc} compiler wrapper to build \mts.  This is installed as part of
the MPI implementation; note that when installing MPI via binary packages,
the development package (if separate from the runtime
support) will likely be needed to compile.
\mts should work on any SMP workstation, properly-configured cluster or general-purpose
supercomputer.  Configuring MPI for clusters is beyond the scope of
this document and depends on the implementation chosen.

\subsection{\mts options}
\label{options}

Similar to the {\em bts\_options} array shown in 
Section~\ref{subsec:btopphase2}, the built-in \mts options are collected in
an {\em mts\_options} array in {\em mts.c}:
{\footnotesize
\begin{lstlisting}
const mtsoption mts_options[] = {       /* external linkage for */
        {"-maxd", 1},                   /* curious applications */
        {"-maxnodes", 1},
        {"-scale", 1},
        {"-lmin", 1},
        {"-lmax", 1},
        {"-maxbuf", 1},
        {"-freq", 1},
        {"-hist", 1},
        {"-checkp", 1},
        {"-stop", 1},
        {"-restart", 1}

};
const int mts_options_size = 11;
\end{lstlisting}
}
\noindent which currently has eleven built-in options, which are similar to options
supported by \mplrs~\cite{AJ15b}. The first option {\em -maxd} takes a parameter that
specifies \mymaxdepth, and likewise {\em -maxnodes} takes a parameter
specifying \maxnodes.  The \maxnodes parameter is scaled by the scaling
factor given by the {\em -scale} option.  The {\em -lmin} and {\em -lmax}
parameters specify when the budget is modified based on the number of unexplored
subtrees available.  The idea is to use smaller budgets to quickly grow the 
job list when it is small, and larger budgets to avoid overhead when the job
list is sufficiently large.

More precisely, the \mymaxdepth is used only if the
number of unexplored subtrees available ($|L|$ in~\cite{AJ15b}) is less than
the number of processes in use times the \lmin value.  The $\maxnodes$ value
is scaled by the scaling factor if the number of unexplored subtrees available
is larger than the number of processes times the \lmax value.

While these parameters have default values used in this paper, it is very
likely that different applications will require different values
to attain good performance.  The next two parameters, {\em -hist} and
{\em -freq} allow one to specify \emph{histogram} and \emph{frequency} files.
These collect statistics on the degree of parallelization obtained, and can
easily be plotted using \gnuplot.  This is intended to help the user tune
\mts to their application.  We explain the format and usage of these files
in Section~\ref{hist}.

The last three options relate to checkpoint files and restarts. 
The {\em -checkp} file is used to specify a \emph{checkpoint} file.  If the run
is interrupted (e.g., by the application using \emph{cleanstop()} to request
a checkpoint), \mts
will save the current list of unexplored nodes, shared data, and other
information needed to restart. The {\em -restart} option is used to
specify a checkpoint file to restart from.  The {\em -stop} option is
used to specify a \emph{stop} file.  When this option is used, \mts will
periodically check for the existence of the specified file.  If it exists,
\mts will checkpoint and exit when possible.

Finally, the option {\em -maxbuf} controls the ``streaminess''
of the output, where the parameter
specifies a number of bytes.  If the buffer is larger than the specified
number of bytes and ends in a newline, it is flushed if possible.  Smaller
values increase streaminess but may increase overhead.
Currently the default parameters are: 
\[
\mymaxdepth=2~~ \maxnodes=5000~~ \myscale=40~~ \lmin=1~~ \lmax=3~~ \maxbuf=1048576.
\]

\section{Computational results}

The tests were performed at Kyoto University on \maic, a cluster of 5 nodes 
with a total of 192 identical processor cores, consisting of: 
\begin{itemize}
\item
\mainew: 4 nodes, each containing: 2x Opteron 6376 (16-core 2.3GHz), 32GB memory, 500GB hard drive (128 cores in total)
\item
\mainewer: 4x Opteron 6376 (16-core 2.3GHz), 64 cores, 256GB memory, 4TB hard drive.
\end{itemize}

\subsection{Linear extensions: \mtop}
\label{subsec:mtop}

The tests were performed using the following codes:
\begin{itemize}
\item
\vr: obtained from the Combinatorial Object Server~\cite{COS}, generates linear 
extensions in lexicographic order via the     
Varol-Rotem algorithm~\cite{VR81} (Algorithm V in Section 7.2.1.2 of~\cite{knuth11})
\item
\genle: also obtained from~\cite{COS}, generates linear extensions in Gray code order
using the algorithm of Pruesse and Rotem~\cite{PR91}
\item
\btop: derived from the reverse search code {\em topsorts.c}~\cite{tutorial} as 
described in Section~\ref{sec:mts}
\item
\mtop: \mts parallelization of \btop.
\end{itemize}
For the tests all codes were used in count-only mode due to the enormous output that 
would otherwise be generated. All codes were used with default parameters given at the end of
Section~\ref{options}.

The problems chosen were the following graphs which are listed in order of increasing
edge density:
\begin{itemize}
\item
{\em pm22} : the partial order $a_1 < a_2,\ldots, a_{21} < a_{22}$,
$a_1 < a_3 <\ldots< a_{21}$ that generates all perfect matchings on 22 elements
\item
{\em cat42} : the partial order $a_1 < a_2,\ldots, a_{41} < a_{42}$;
$~~~a_1 < a_3,\ldots, a_{39} < a_{41}$; $~~~a_2 < a_4,\ldots, a_{40} < a_{42}$
that generates all the 2 x 21 Young Tableaux, whose cardinality is the
21st Catalan number  
\item
$K_{8,9}$ : the partial order $a_1 < a_9,\ldots, a_1 < a_{17}$
$~~ldots~~$ $a_8 < a_9,\ldots, a_8 < a_{17}$ corresponding to 
the complete bipartite graph $K_{8,9}$
with all edges directed from the smaller part to the larger.
\end{itemize}
The constructions for the first two partial orders
are well known (see, e.g., Section 7.2.1.2 of~\cite{knuth11}).

\begin{table}[htbp]
\centering
\scalebox{0.88}{
\begin{tabular}{||c c c c||c|c|c||c|c|c|c|c||}
 \hline

Graph &m&n   & No. of perms  &\vr& \genle  & \btop  &\multicolumn{5}{|c||}{\mtop  }   \\
      & &    &   & &   &  & 12 & 24 & 48 & 96 & 192  \\
\hline
{\em pm22} &22&21    & 13,749,310,575  & 179  & 14 & 12723  & 1172 &595& 360&206 &125 \\
 & &  &   & &  & (1) & (11)   & (21)    & (35)    & (62)     & (102)    \\
\hline
{\em cat42} &42 &61   & 24,466,267,020  & 654  & 171 &45674  &4731 &2699&1293&724 &408 \\
 & &  &   & &  & (1) & (9.7)   & (17) & (35)  & (63)  & (112)    \\
\hline
$K_{8,9}$ &17&72    & 14,631,321,600  & 159  & 5 & 8957 &859&445&249 & 137   &85  \\
 & &  &   & &  & (1) & (10)   & (20)    & (36)    & (65)  & (105)    \\
\hline
\end{tabular}
}
\caption{Linear extensions: \maic, times in secs (speedup on \btop)}
\label{tab:tops}
\end{table}

The computational results are given in Table~\ref{tab:tops}. First observe that the reverse search
code \btop is very slow, over 900 times slower than \genle and
over 70 times slower than \vr on {\em pm22} for example. However     
the parallel \mts code obtains excellent speedups and is faster than \vr on all problems when
192 cores are used. 

\subsection{Spanning trees: \mtree}

The tests were performed using the following codes:
\begin{itemize}
\item
\gray: Knuth's implementation~\cite{knuthcode} of an algorithm that generates all spanning trees of a given graph,
changing only one edge at a time, as described in
Malcolm Smith's M.S. thesis, {\em Generating spanning trees} (University
of Victoria, 1997)
\item
\graysp: Knuth's improved implementation of \gray:
``This program combines the ideas of \gray
and {\em spspan}, resulting in a glorious routine that generates
all spanning trees of a given graph, changing only one edge at a time,
with `guaranteed efficiency'---in the sense that the total running
time is $O(m+n+t)$ when there are $m$ edges, $n$ vertices, and $t$
spanning trees.''\cite{knuthcode}
\item
\btree: derived from the reverse search code {\em tree.c}~\cite{tutorial} as described in Section~\ref{sec:tree}
\item
\mtree: \mts parallelization of \btree.
\end{itemize}
\noindent
Both \gray and \graysp are described in detail in Knuth~\cite{knuth11}.
Again all codes were used in count-only mode 
and with the default parameters given at the end of Section~\ref{options}.

The problems chosen were the following graphs which are listed in order of increasing
edge density:
\begin{itemize}
\item
{\em 8-cage} : the Tutte-Coxeter graph, which is the unique minimal graph of girth 8
\item
$P_5C_5$ : 5X5 cylinder $P_5 \Box C_5$
\item
$C_5C_5$ : 5X5 cylinder $C_5 \Box C_5$
\item
$K_{7,7}$ : the complete bipartite graph with 7 vertices in each part
\item
$K_{12}$ : the complete graph on 12 vertices.
\end{itemize}
The latter 4 graphs were motivated by Table 5 in~\cite{knuth11}: $P_5C_5$ appears therein
and the other graphs are larger versions of examples in the table.

\label{sec:comp}
\begin{table}[h!tbp]
\centering
\scalebox{0.9}{
\begin{tabular}{||c c c c||c|c|c||c|c|c|c|c||}
 \hline

Graph &m&n   & No. of trees  &\gray& \graysp & \btree &\multicolumn{5}{|c||}{\mtree  }   \\
 & &    &   & &  &  & 12 & 24 & 48 & 96 & 192  \\
\hline
{\em 8-cage} &30&45    &  23,066,015,625   & 3166 & 730 & 10008 &1061 &459& 238&137 & 92  \\
 & &  &   & &  & (1) & (9.4)    & (21)    & (42)    & (73)     & (109)    \\
\hline
$P_5C_5$  & 25& 45 & 38,720,000,000  & 3962 & 1212& 8918  & 851 &455& 221&137 & 122 \\
 & &  &   & &  & (1) & (10)   & (20)    & (40)    & (65)     & (73)    \\
\hline
$C_5C_5$  & 25 & 50 &1,562,500,000,000  & 131092 & 41568&230077    &26790 &13280&7459& 4960 & 4244 \\
 & &  &   & &  & (1) & (8.6)   & (17)    & (31)  & (46)     & (54)    \\
\hline
$ K_{7,7}$ & 14 & 49 &  13,841,287,201   &  699 & 460 &  2708 &259  &142&  68& 51 & 61 \\
 & &  &   & &  & (1) & (10)   & (19)    & (40)    & (53)     & (44)    \\
\hline
$K_{12}$ & 12 & 66 &  61,917,364,224   & 2394 & 1978 & 3179 &310 &172&  84& 97 & 148 \\
 & &  &   & &  & (1) & (10)   & (18)    & (38)    & (33)     & (21)    \\
\hline
\end{tabular}
}
\caption{Spanning tree generation: \maic, times in secs (speedup on \btree)}
\label{tab:trees}
\end{table}
The computational results are given in Table~\ref{tab:trees}. This time the reverse search
code is a bit more competitive:  about 3 times slower than \gray and
about 14 times slower than \graysp on {\em 8-cage} for example. 
The parallel \mts code runs about as fast as \graysp on all problems when
12 cores are used and is significantly faster after that. Near linear speedups are
obtained up to 48-cores but then tail off. For the two dense graphs
$ K_{7,7}$ and $K_{12}$ the performance of \mts is actually worse with 192 cores than with 96. 

\section{Evaluating and improving performance}
\label{hist}

The amount of work contained in a node can vary dramatically between
applications, and so \mts includes features intended to help tune
its budgeting to a particular application.  We briefly mentioned
histogram and frequency files in Section~\ref{subsec:mtsopts};
here we take a closer look at using these to improve performance.
We focus on \mtop and the $K_{8,9}$ instance introduced in
Section~\ref{subsec:mtop}.

When \mts is used with the \emph{-hist} option, it periodically writes
a line to the specified histogram file.  The line contains (in order
and separated by whitespace):
\begin{enumerate}
 \item Time in seconds since execution began (floating point)
 \item The number of workers currently busy
 \item The number of unexplored nodes currently in the joblist
 \item The number of workers owing the master a message about
       unexplored nodes
 \item Unused (currently $0$)
 \item Unused (currently $0$)
 \item The total number of unexplored nodes (jobs) the master has seen
       since execution began.
\end{enumerate}
For example, the following is part of a histogram file generated by
\mtop using 128 cores and default parameters:
\begin{verbatim}
1.003161 114 1780 114 0 0 6297
2.005773 120 1675 120 0 0 9946
3.006864 117 1679 117 0 0 13665
\end{verbatim}
This indicates that at (approximately) 3 seconds, there were
117 busy workers, 1679 jobs in the list, 117 workers owing messages, and
13665 total jobs had existed (i.e. completed jobs, currently executing 
jobs, and the jobs currently in the list).  

The number of workers owing messages to the master is always at least
the number of busy workers (since any busy worker must report the number
of unexplored nodes it will return).  It could become larger, for example
 if these messages are being delayed due to a saturated interconnect.
However, the other entries are likely to be of more interest.

\begin{figure}[h!tbp]
\begin{minipage}{0.5\textwidth}
 \includegraphics[width=\textwidth]{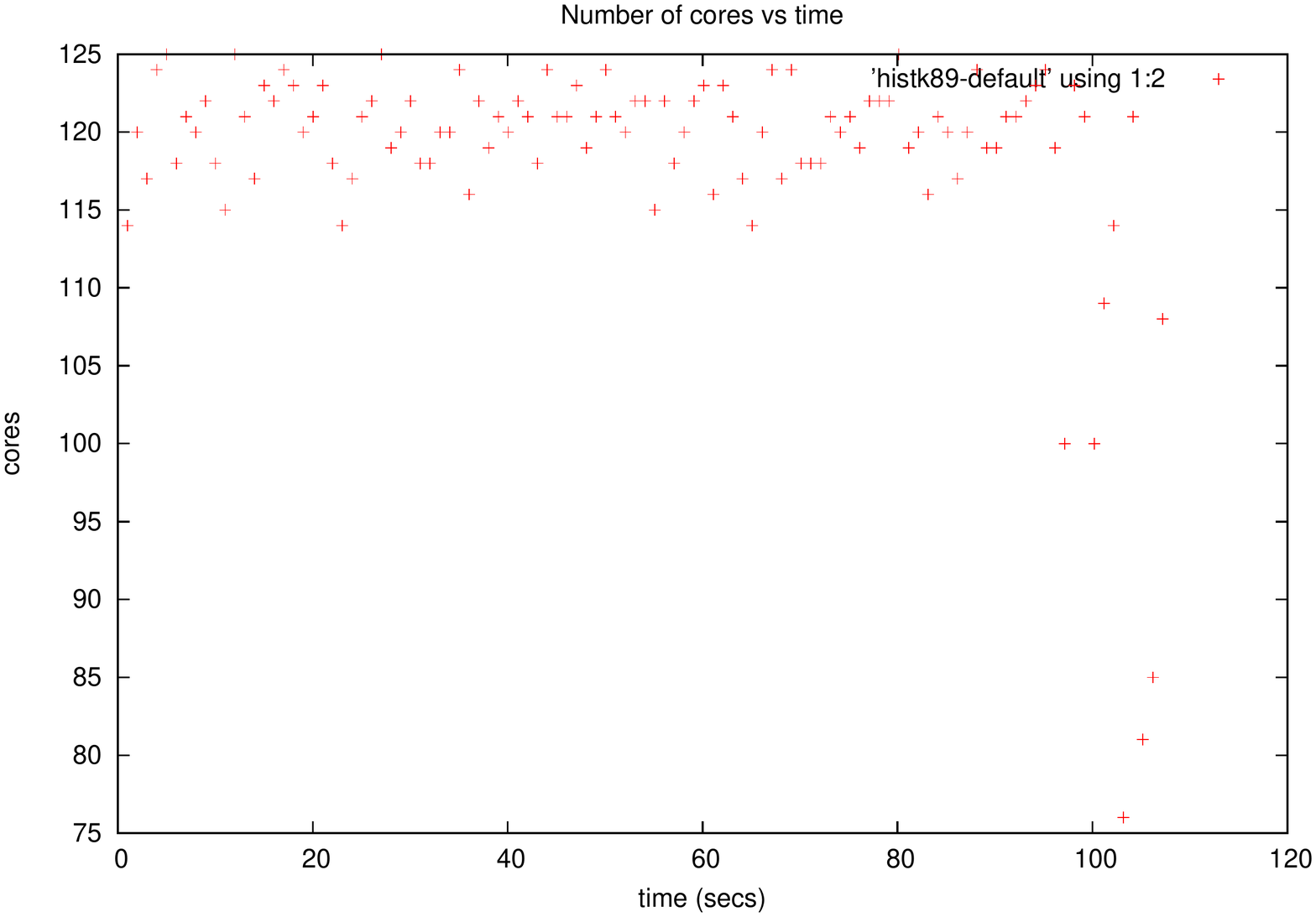}
\end{minipage}
\begin{minipage}{0.5\textwidth}
 \includegraphics[width=\textwidth]{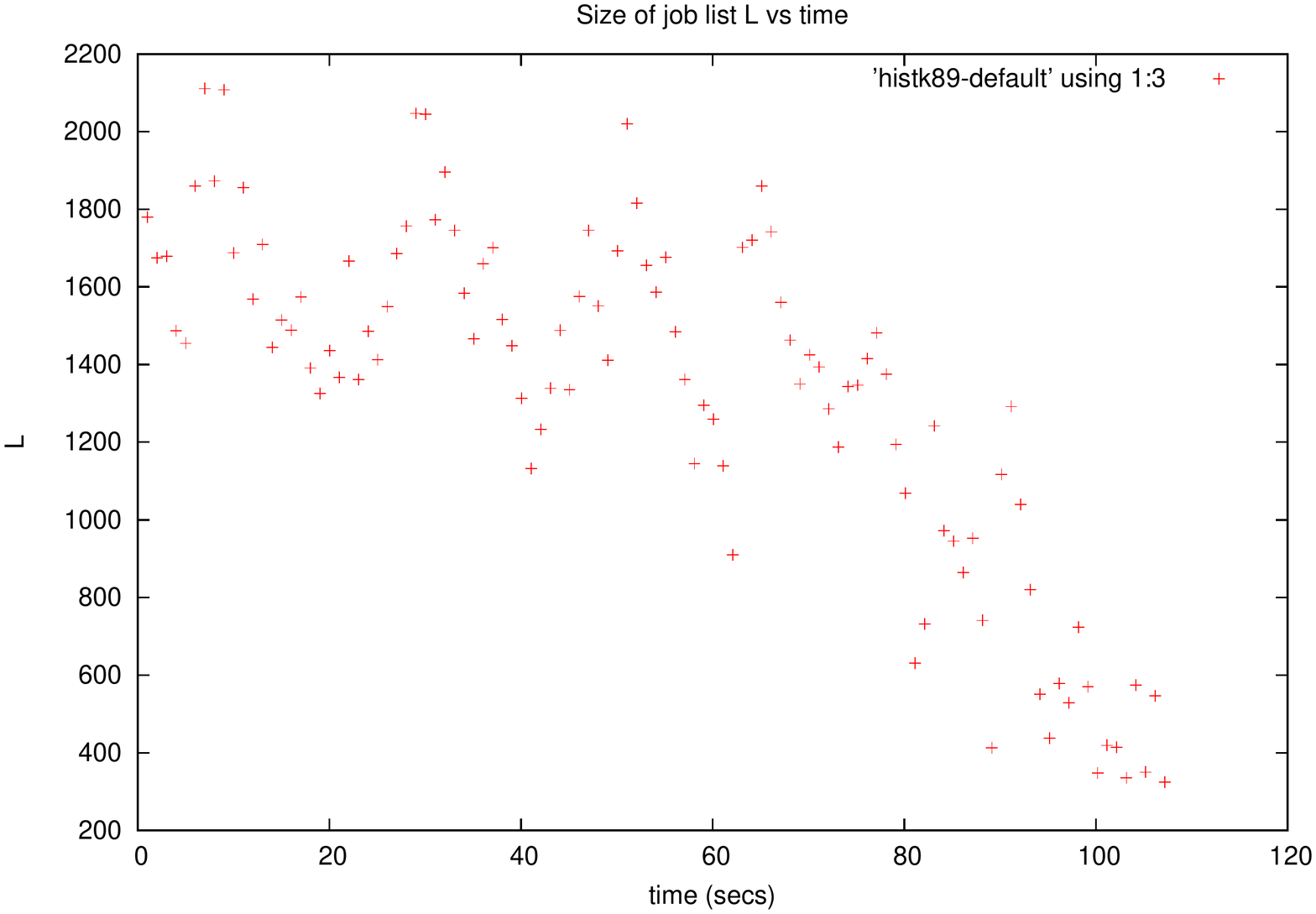}
\end{minipage}
\begin{minipage}{0.5\textwidth}
 \includegraphics[width=\textwidth]{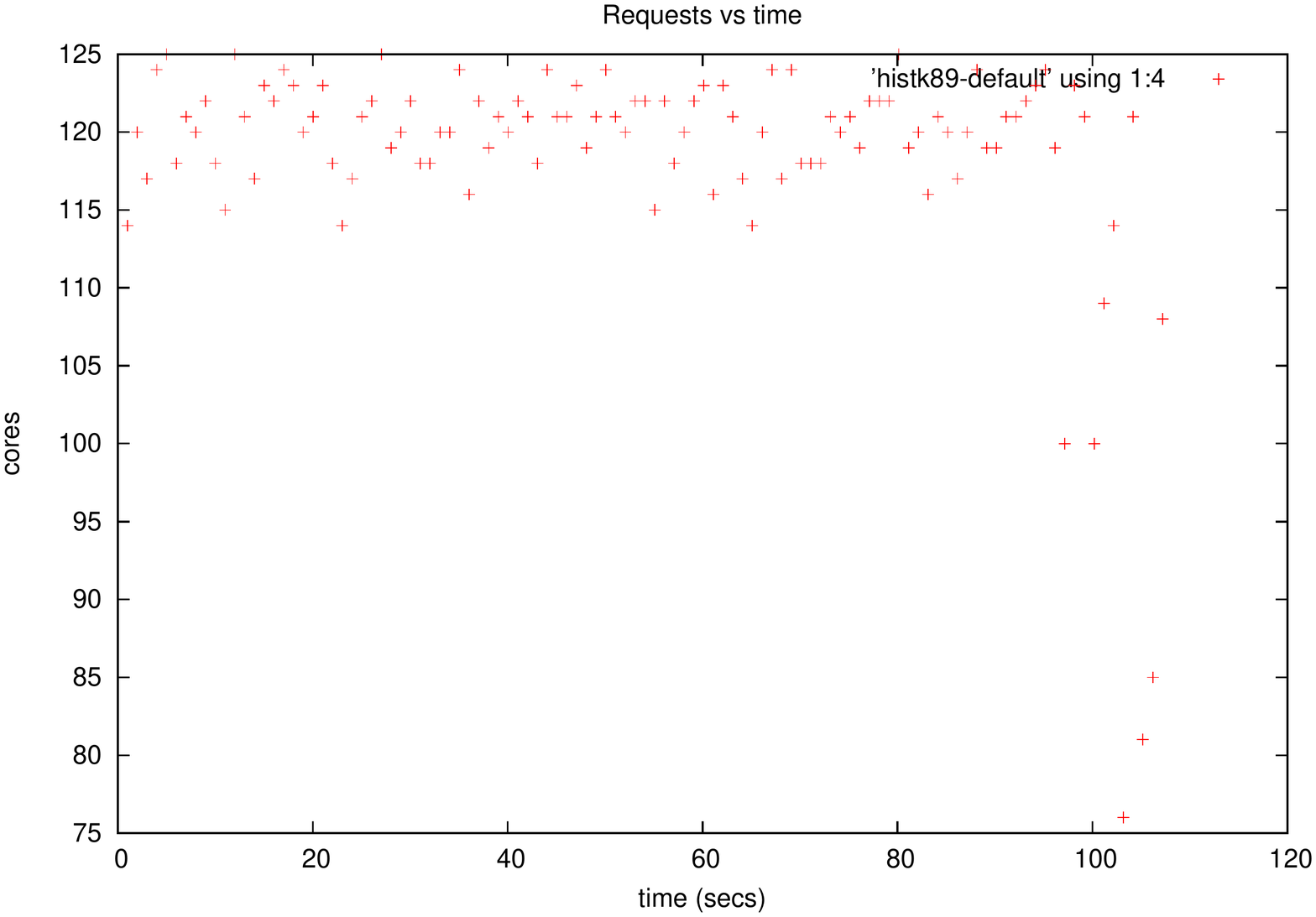}
\end{minipage}
\caption{Histograms for \mtop on $K_{8,9}$: busy workers, job list size, messages owed }
\label{fig:defplot}
\end{figure}

The \mts distribution includes a script (\texttt{plotL.gp}) to produce
graphical plots from these histogram files using \gnuplot.  The script
produces three plots, with time on the horizontal axis and the
second to fourth columns on the vertical axis.  Figure~\ref{fig:defplot}
contains the plots produced for the \mtop run with default parameters 
in the example above.

In Figure~\ref{fig:defplot} using the default parameters, we see that the master is struggling to keep 
all workers busy despite having jobs available.  This suggests that we can 
improve performance by using better parameters.
In this case, a larger \emph{-scale} or \emph{-maxnodes} value may help,
since it will allow workers to do more work (assuming a sufficiently large 
subproblem) before contacting the master.

\begin{figure}[h!tbp]
\centering
\begin{minipage}{0.49\textwidth}
 \includegraphics[width=\textwidth]{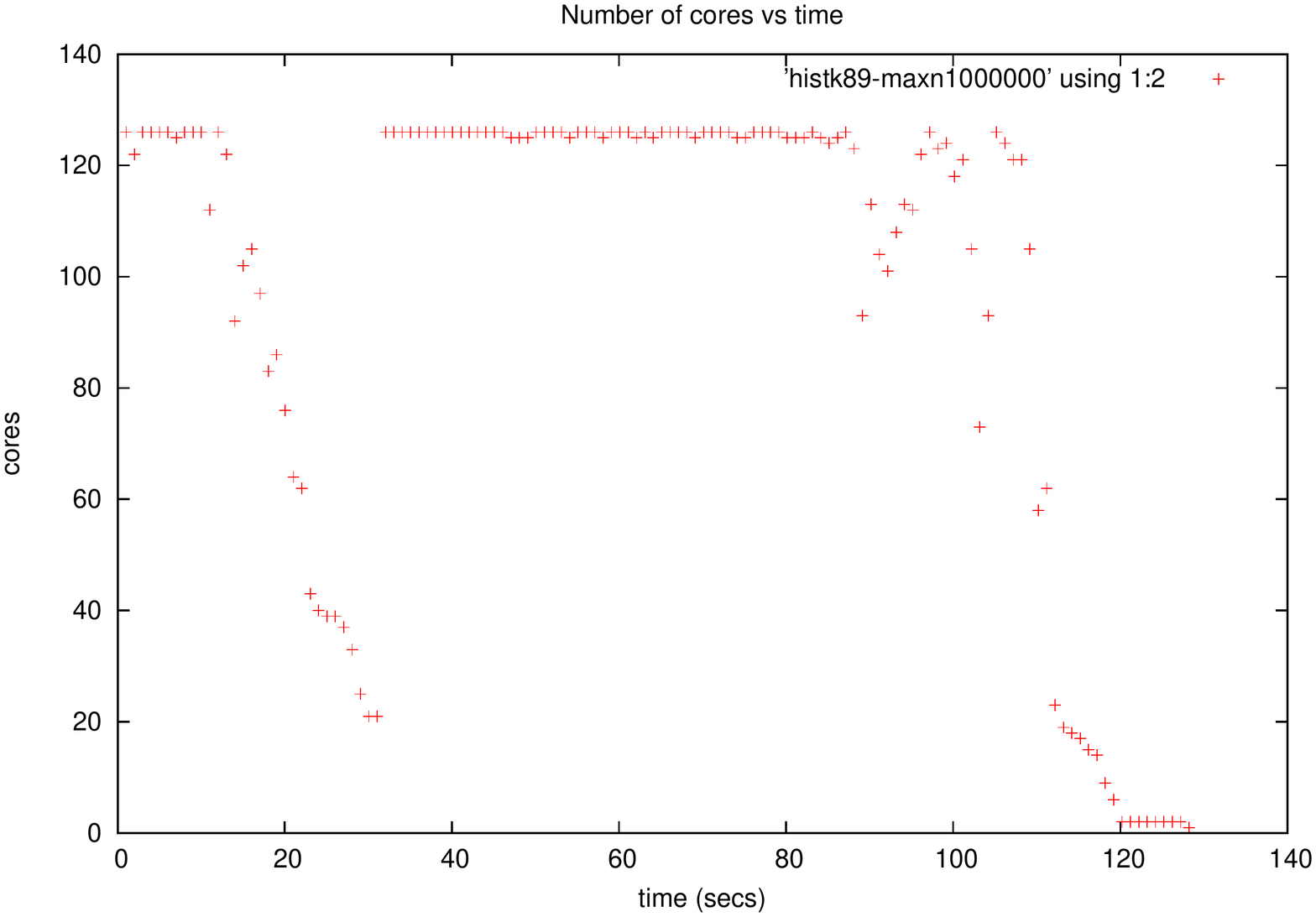}
\end{minipage}
\begin{minipage}{0.49\textwidth}
 \includegraphics[width=\textwidth]{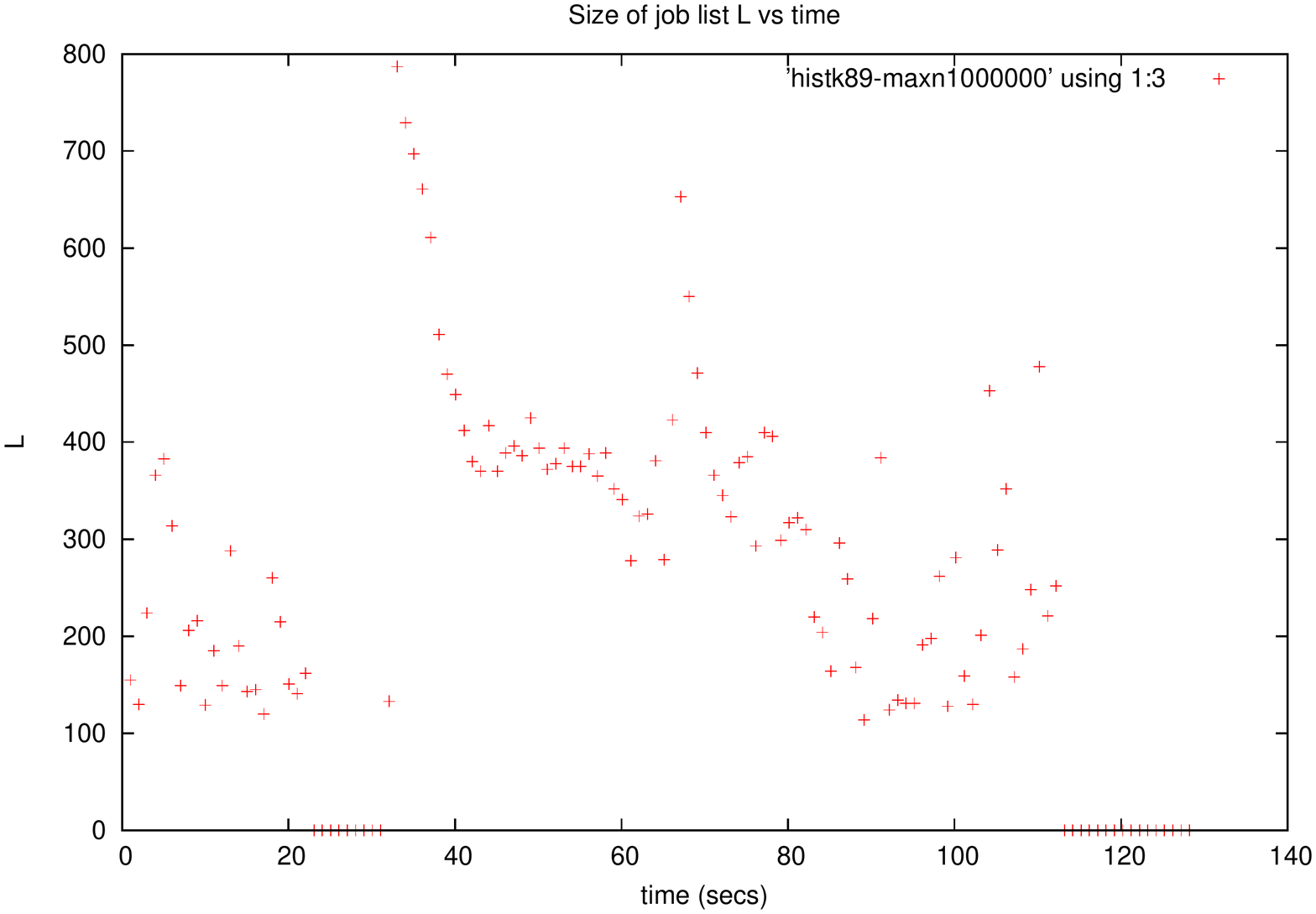}
\end{minipage}
\caption{Histograms with \emph{-maxnodes} $1000000$: 
busy workers (l), joblist size (r)}
\label{fig:maxnplot}
\end{figure}

For example, we might try increasing \emph{-maxnodes}.  
Figure~\ref{fig:maxnplot} shows\footnote{We omit the third plot; the number
of messages owed normally tracks the number of busy workers.}
the result of using the (much larger) value $1000000$.  There
we see the opposite problem; many workers become idle
because the job list becomes empty.  This suggests that \emph{-maxnodes} 
$1000000$ is not suitable in this case; even when the job list is nearly
empty the master schedules jobs with large budgets.

\begin{figure}[h!tbp]
\centering
\begin{minipage}{0.49\textwidth}
 \includegraphics[width=\textwidth]{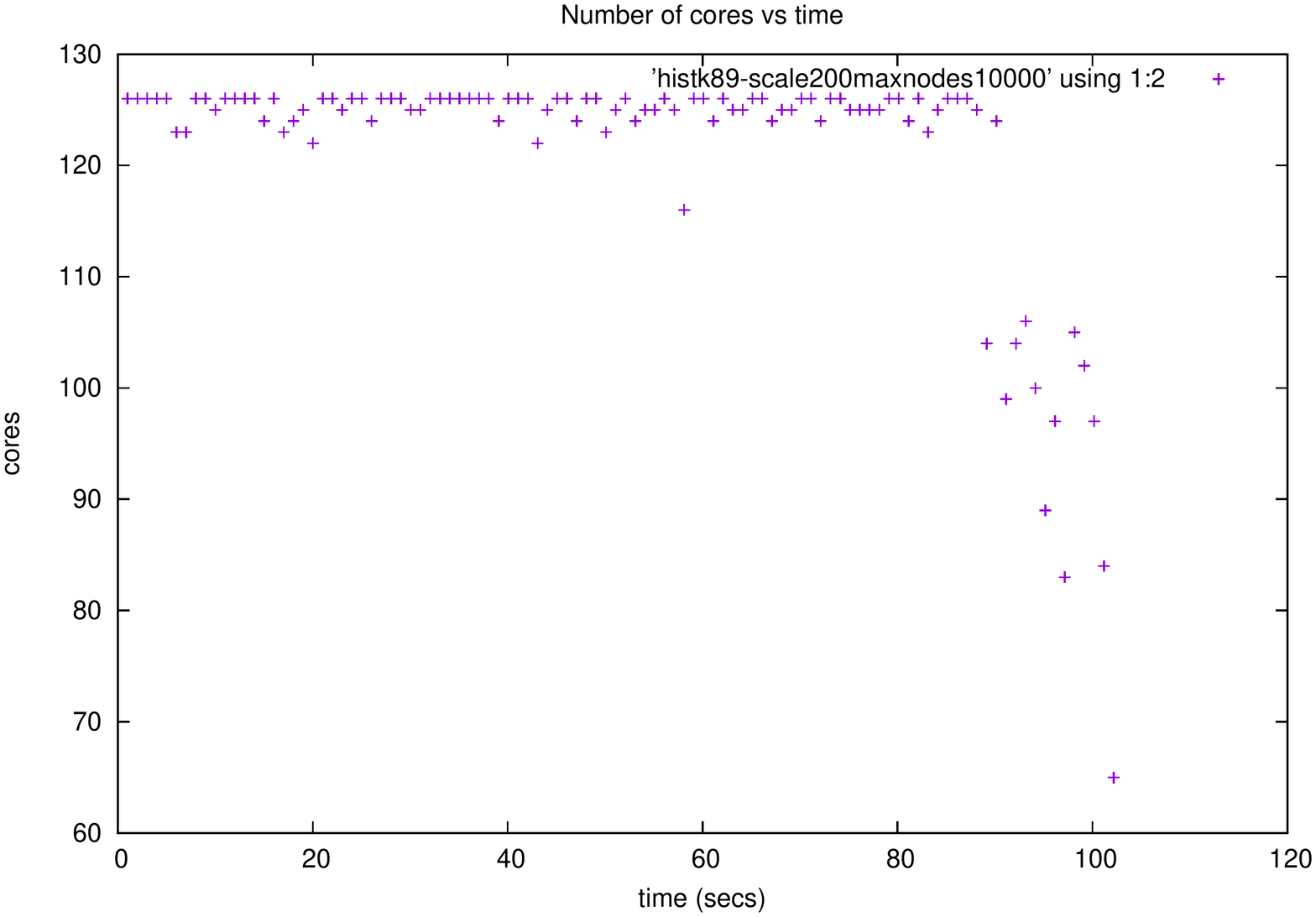}
\end{minipage}
\begin{minipage}{0.49\textwidth}
 \includegraphics[width=\textwidth]{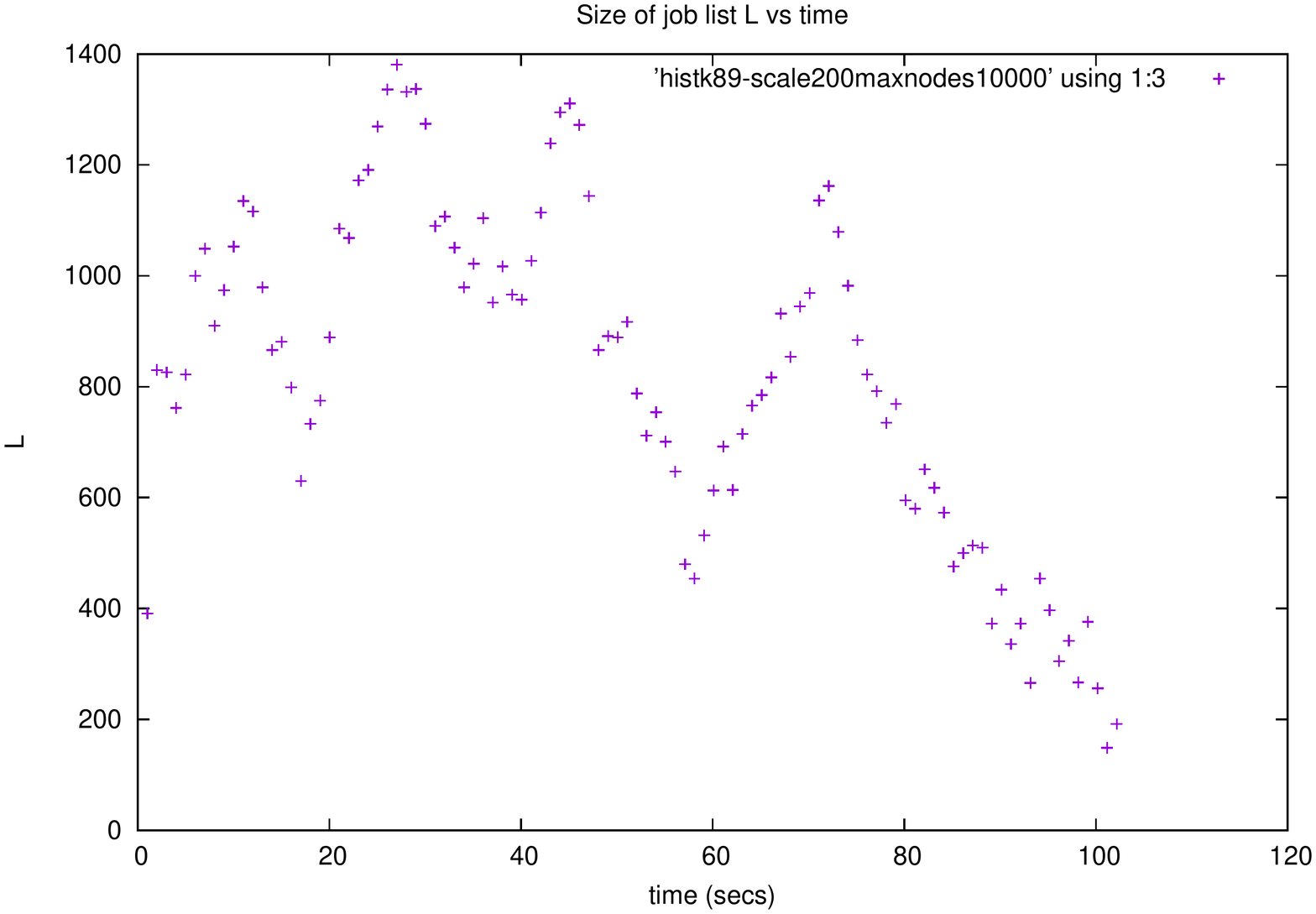}
\end{minipage}
\caption{Histograms with \emph{-scale} $200$ \emph{-maxnodes}
$10000$ on $K_{8,9}$: 
busy workers (l), joblist size (r)}
\label{fig:scaleplot}
\end{figure}

Instead of dramatically increasing \emph{-maxnodes}, we might try a 
larger \emph{-scale} together with a modest increase in \emph{-maxnodes}.
This will leave budgets relatively low when the job list is small,
and increase the budget by a larger amount when the job list is large.
Figure~\ref{fig:scaleplot} shows the result of using a value of $200$ for
scaling and $10000$ for \emph{-maxnodes}.  These parameters produce less
than half the number of total number of jobs compared to the default
parameters, and increase overall performance by about five
percent on this particular input. 

In addition to the performance histograms, \mts can generate
\emph{frequency} files.  These files contain one number per line,
giving the values that the application's \emph{bts()} function
returns on each job.  In our sample applications, this value is
the number of nodes that \emph{bts()} visited -- and so the
frequency file contains the actual size of each job.  This can
provide statistical information about the tree that is helpful
when tuning the parameters for better performance.  For example,
it may be particularly helpful to implement and use pruning if many
jobs correspond to leaves.  Likewise, increasing the budget will
have limited effect if only few jobs use the full budget.

\begin{figure}[h!tbp]
 \begin{minipage}{0.49\textwidth}
  \includegraphics[width=\textwidth]{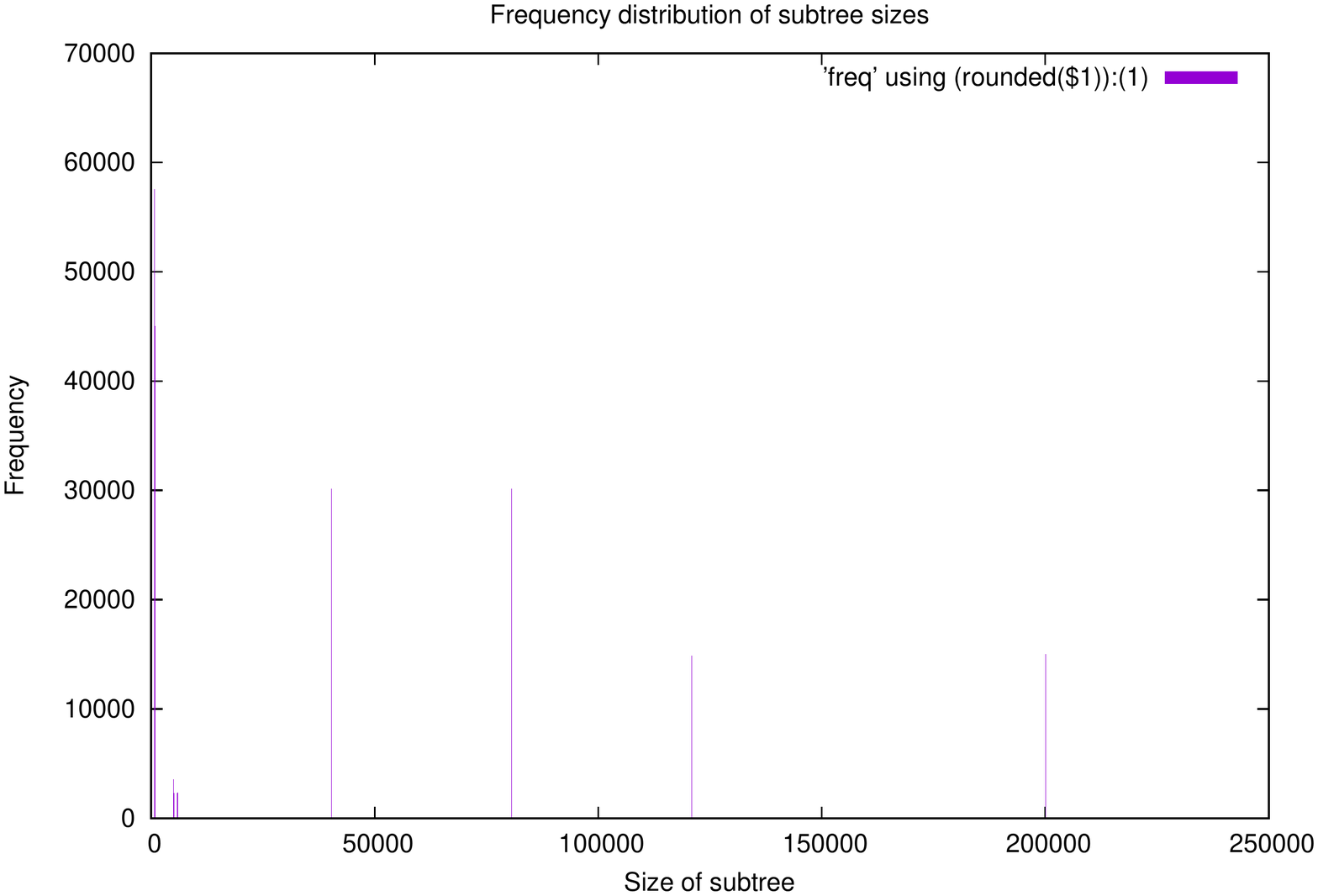}
 \end{minipage}
 \begin{minipage}{0.49\textwidth}
  \includegraphics[width=\textwidth]{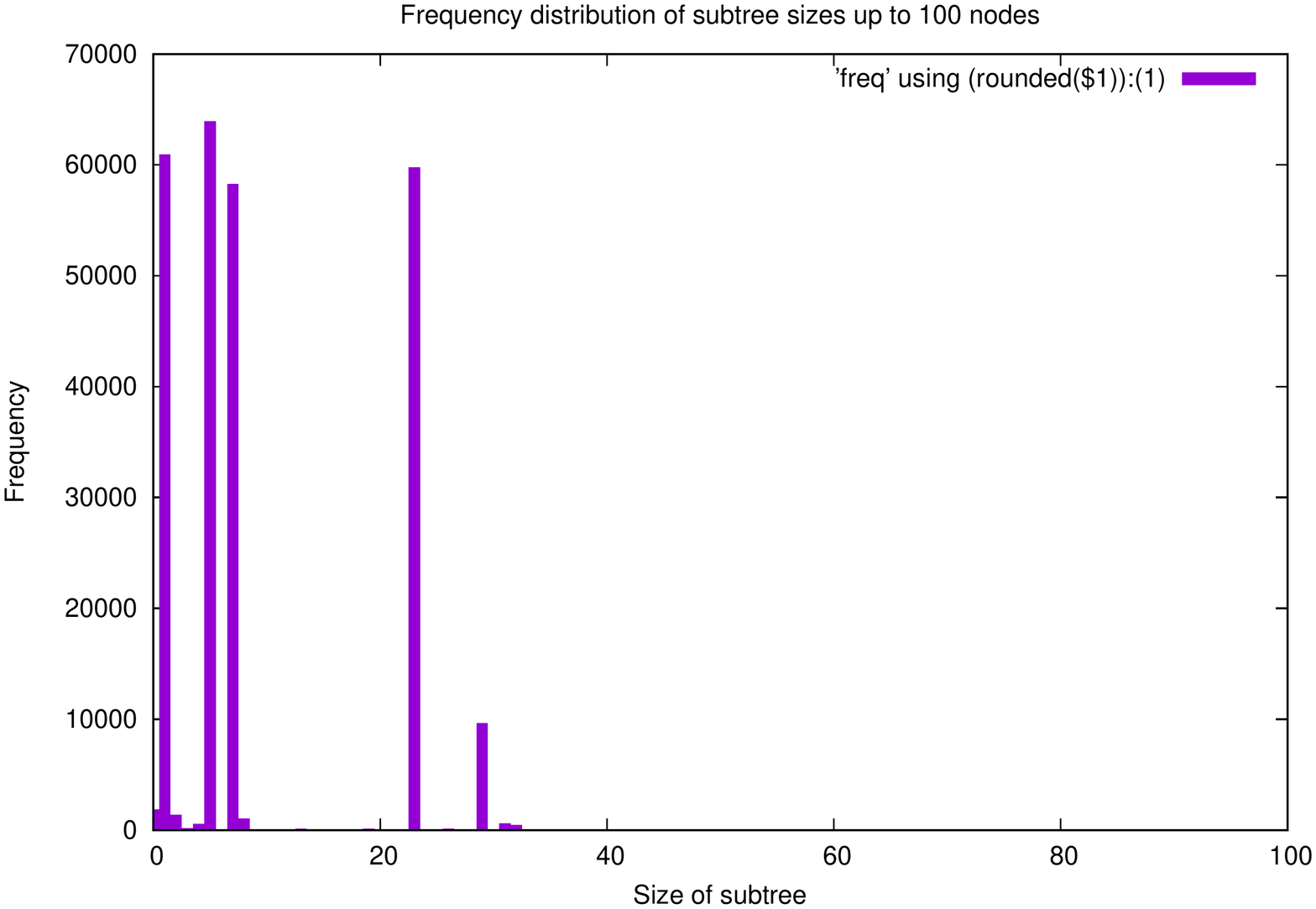}
 \end{minipage}
 \caption{Subproblem sizes in $K_{8,9}$, default parameters: all (l) small subproblems only (r)}
 \label{fig:freqplot}
\end{figure}

The \mts distribution contains a script (\texttt{plotD.gp}) that
uses \gnuplot to plot the distribution of subproblem sizes.
Figure~\ref{fig:freqplot} shows the distribution of subproblem
sizes that was produced in a run of \mtop on $K_{8,9}$ with
default parameters.

\section{Conclusions}
We have presented a generic framework for parallelizing reverse search codes requiring only
minimal changes to the legacy code. Two features of our approach are that the parallelizing wrapper
does not need to be user modified and the modified legacy code can be tested in standalone
single processor mode. Applying this framework to
two very basic reverse search codes we obtained comparable results
to that previously obtained by the customized \mplrs wrapper applied to the
the complex \lrs code~\cite{AJ15b}. We expect that many other 
reverse search applications will obtain similar speedups when parallelized with \mts.

Ongoing work is to extend the use of \mts to more complex search tree algorithms
such as branch and bound, SAT solvers, game tree search, etc. Since these applications require
some sharing of common data it will be interesting to see if the same 
sorts of speedups can be obtained.
\bibliographystyle{spmpsci}
\bibliography{tutorial2}

\begin{thebibliography}{1}
\providecommand{\url}[1]{{#1}}
\providecommand{\urlprefix}{URL }
\expandafter\ifx\csname urlstyle\endcsname\relax
  \providecommand{\doi}[1]{DOI~\discretionary{}{}{}#1}\else
  \providecommand{\doi}{DOI~\discretionary{}{}{}\begingroup
  \urlstyle{rm}\Url}\fi

\bibitem{COS}
Combinatorial {O}bject {S}erver : Linear extensions.
\newblock \url{http://theory.cs.uvic.ca/inf/pose/LinearExt.html}

\bibitem{tutorial}
Avis, D.: Tutorial on reverse search with {C} implementations (2000).
\newblock \url{http://cgm.cs.mcgill.ca/~avis/doc/tutorial}

\bibitem{AF93}
Avis, D., Fukuda, K.: Reverse search for enumeration.
\newblock Discrete Applied Mathematics \textbf{65}, 21--46 (1993)

\bibitem{AJ15b}
Avis, D., Jordan, C.: mplrs: {A} scalable parallel vertex/facet enumeration
  code.
\newblock CoRR \textbf{abs/1511.06487} (2015).
\newblock \urlprefix\url{http://arxiv.org/abs/1511.06487}

\bibitem{knuthcode}
Knuth, D.E.: Programs to read.
\newblock \url{http://www-cs-faculty.stanford.edu/~uno/programs.html}

\bibitem{knuth11}
Knuth, D.E.: The Art of Computer Programming, Volume 4A.
\newblock Addison-Wesley Professional (2011)

\bibitem{PR91}
Pruesse, G., Ruskey, F.: Generating the linear extensions of certain posets by
  transpositions.
\newblock SIAM Journal on Discrete Mathematics \textbf{4}(3), 413--422 (1991)

\bibitem{VR81}
Varol, Y.L., Rotem, D.: An algorithm to generate all topological sorting
  arrangements.
\newblock The Computer Journal \textbf{24}(1), 83--84 (1981)

\end{thebibliography}

\end{document}